\documentclass[twocolumn,showpacs, showkeys,preprintnumbers,amsmath,amssymb]{revtex4}

\usepackage{graphicx}
\usepackage{dcolumn}
\usepackage{bm}

\begin{document}

\preprint{draft}

\title{A student designed experiment measuring the speed of sound as a function of altitude} 

\author{G. Santostasi$^{a}$ and MESS team: D. Hughes$^{a}$, P. Maharjan$^{c}$, C. McAdon,  N. T. Nguyesn$^{b}$, S. Poudel$^{b}$, S. Pradhan$^{b}$, D. Roshan$^{b}$, and M. Wagle$^{c}$}

\affiliation{ \footnotesize{ $^{a}$Physics Department\\
\footnotesize $^{b}$Engineering Department\\
\footnotesize $ ^{c}$ Mathematics, Computer Science and Statistics Department\\ McNeese State University}
}%

\date{\today}
             
\begin{abstract}
\begin{center}
\bf{ \normalsize{Abstract} }
\end{center}
Relatively inexpensive and readily commercially available equipment (such as digital recorders, MP3 portable speakers and tie-pin microphones), allowed a team of students from McNeese State University to measure the speed of sound in the atmosphere as a function of altitude. The experiment was carried as a payload (in the context of a NASA funded student program called La-ACES) on a high altitude balloon that reached a maximum altitude of 101,000 feet. Not withstanding substantial environmental noise, our particular experimental design allowed for the filtering of the signal out of the noise, thus achieving remarkable accuracy and precision. The speed of sound measurement was then used to set limits on the abundances of the main molecular components of the atmosphere (diatomic nitrogen and oxygen). Bayesian analysis was used to set meaningful values on the uncertainty of our limits. It is our experience that students find intutive and appealing this type of probability method.

\end{abstract}
                            
\maketitle

\section{Introduction}  
La-ACES (Louisiana Aerospace Catalyst Experience for Students) \cite{c7} is an ongoing NASA funded project sponsored by the Louisiana Space Consortium. Students are recruited to form teams from different Louisiana universities. The teams receive training (on various aerospace subjects) over a semester and spend another semester working on a payload that will fly on a high altitude balloon. The payloads are designed by the students with minimal help and advice by a supervisor that is usually an instructor at the students' institution.\\ McNeese State University's students, participating in the 2008 LA-ACES program, have opted to measure the speed of sound in the atmosphere. They have achieved this goal with remarkable accuracy and precision considering LA-ACES is a student project. We present, in this article, data and methods from a recent balloon launch (end of May 2008) taken place at the NASA Columbia Scientific Balloon Facility in Palestine, Tx.\\
In recent years, educators have proposed different designs for experimental setups taking advantage of the new technology of various digital devices. Measurements of the speed of sound, using computer sound cards and audio manipulation software, have been described in the literature, see \cite{c4}, \cite{c6} and references therein. In terms of the availability of components, portability, light weight (about 1 kg), and simplicity we believe our design is quite unique. Our speed of sound device is also very accurate and precise even in the presence of substantial environmental noise.
The light weight of our apparatus allows for experiments outside the laboratory. In particular, we have had success with students' exploration of the physical properties of the atmosphere when our device was used as payload on a high altitude balloon. The availability of weather balloons and their relatively inexpensive cost makes them very accessible to colleges and universities. High altitude weather balloons offer platforms to perform very interesting experiments that can illustrate different principles that would not be so fascinating to the students if performed only in the laboratory.
Our experience is that students are excited in being involved in such exploration of near space, often demonstrating a sense of ownership of the project, and they are really eager to use different analysis tools to extract any useful information from the data.\\
We have used our speed of sound experiment to introduce students to Bayesian data analysis methods that are becoming more and more prominent in the scientific community. The students seem to be able to grasp the fundamental ideas of Bayesian probability naturally. We give a short introduction to the main Bayesian concepts and show the analysis that we have used to process our data. Similar experiments can be used as an introduction to this important and current analysis tool.
\section{Sounds from another world}
Titan is the second largest moon in our solar system and the largest of Saturn's moons. Titan also exibits the only other nitrogen rich atmosphere (with a main composition of 98.4 percent nitrogen and the rest methane) in our solar system, besides the Earth. The moon is surrounded by a dense fog.  The fog, or haze, is thought to be made up of hydrocarbons that form from methane in the atmosphere being broken up by ultraviolet light given off by the Sun. The sunlight that isn't absorbed by the formation of hydrocarbons is reflected by the haze which causes Titan to experience very cool temperatures, down to minus 180 degrees Celsius. The characteristic orange haze of Titan also causes the features of the surface of the moon to be, for the most part, unknown. To gain more information on such a mysterious place a landing probe was needed.
The European Space Agency (ESA) launched the Cassini-Huygens mission on October 15, 1997 \cite{c9}. The overall purpose of the mission was to learn more about Saturn, its fascinating rings and moons system. The Huygens probe was designed specifically to study the atmosphere and the surface of Titan.  The Huygens probe separated from Cassini on December 25, 2004, and landed on the surface of the Saturn's larger moon on January 14, 2005. After the probe landed, it continued sending information back to Earth for about 90 minutes. On board of Huygens was a set of microphones to analyze environmental noise as the probe descended through Titan's atmosphere. Huygens was one of the few probes to ever have carried instruments to record the sound of an alien world \cite{c10}. One of the motivations to add a set of microphones to the probe's scientific payload was to listen for any signs of thunder to verify if the moon actually had any lightning. The microphones were also used to determine the speed of sound as a function of altitude (using a time delay technique similar to the one adopted in our students' experiment). The speed of sound measurement in an extraterrestrial atmosphere was an interesting scientific goal in itself but the data was eventually used in a recent article, published in the prestigious planetary science journal Icarus, to determine limits on the chemical composition of Titan's atmosphere \cite{c1}. The Huygens mission, and in particular its measurement of the speed of sound, was a major inspiration for our students.\\The speed of sound depends on several physical parameters, some of which are independent of chemical composition (like the temperature of the gas) and others (like the density and bulk modulus) that are characteristic of the particular chemical composition of the medium. See section VII. Consequently, the speed of sound can be used to determine (if environmental parameters like the temperature of the gas are known) the precise combination of individual molecules present in the medium. A prior knowledge of the possible components of the gas is somehow necessary to determine limits on the relative chemical abundances of the different molecules. Bayesian statistical methods are the most appropriate for this kind of analysis.
We have used such methods (see section VIII) to determine the chemical composition of the Earth's atmosphere, even if it is a well known quantity. This is in the following of the spirit of the LA-ACES program that invites the students' teams to pretend they are exploring the Earth, as a space mission would, when it explores another world.

\begin{figure}
\centering	\includegraphics[width=0.45\textwidth]{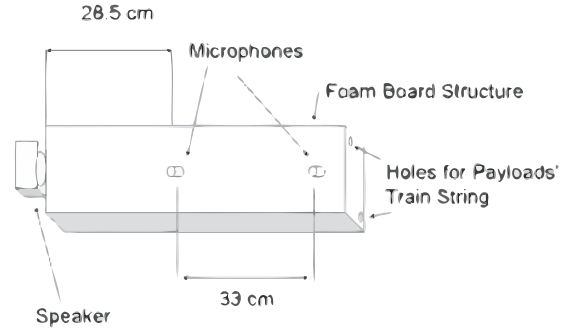}
	\caption{A simple drawing of the external structure of the payload box. The microphones are separated by a distance of 33 cm and the portable speaker is at 28.5 cm from the first microphone. The walls of the box are made of foamboard. The box hosts wiring, two digital recorders and the battery case for the speaker.}
	\label{fig:atmosphere1}
\end{figure}

\begin{figure}
	\centering	\includegraphics[width=0.45\textwidth]{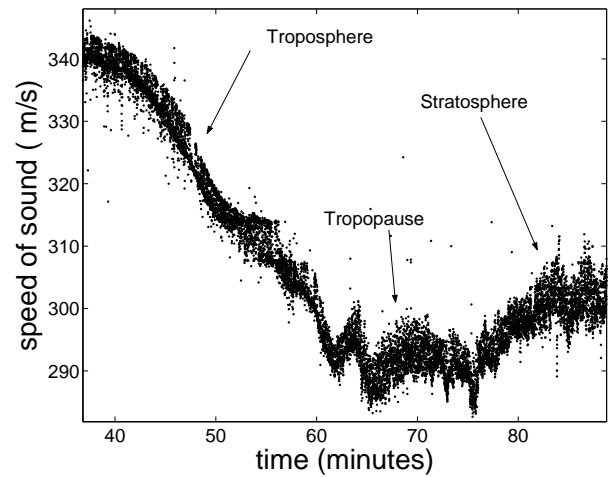}
	\caption{These are the original data points. The data, actually, were acquired at a higher sampling frequency of 96kHz. The speed of sound is determined by measuring the delay time between the peaks of two sine functions that represent the response to a single sound signal of two microphones separated by a distance of 33 cm. The original sound signal is a sine wave created by a speaker at a distance of 28.5 cm from the first microphone. The signal has a precise frequency of 1920 Hz. We have adopted a band pass Chebychev filter (with a bandwidth of about 50 Hz) around the signal to reduce environmental and instrumental noise. The data points shown in the figure are averages (taken over a period of 0.25 seconds) of the high frequency sampled data. The data show a precise trend with a certain spread and also few outliers (that are given to environmental high frequency noise as gushes of wind).}
	\label{fig:original1}
\end{figure}

	\begin{figure}

	\centering	\includegraphics[width=0.45\textwidth]{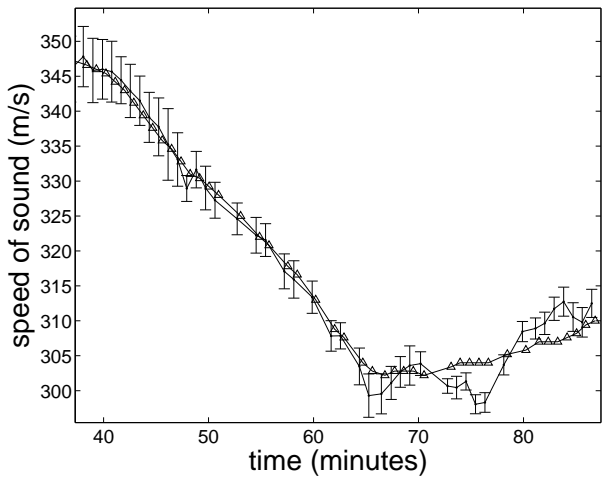}
	\caption{The data after an initial low pass filter that averaged the numerous data points and eliminated the outliers (mostly due to noise caused by gushes of wind). The spread in the data is represented by error-bars (with a width of one standard deviation). Furthermore, the data has been decimated by a factor of 300 to show better the data points and error-bars. Finally, we adjusted the speed of sound to take into account the measured ascension velocity of the balloon. The triangles represent a simple theoretical model of the speed of sound as a function only of the temperature (the model implies the medium to be an ideal gas with the molecular composition of air at sea level). Temperature data were acquired by an external sensor in the GPS transmitter box, at the end of the payload line (few meters away from our payload). The fitting is remarkably good (notice that the measured range in the speed of sound data is only about 10 percent). For a more quantitatively analysis of the goodness of the fit, between experimental and theoretical data, see Fig. 4.}
	\label{fig:fittedtheory1}
\end{figure}
.  
\section{Payload Design}
The payload consisted of a few main components: the external structural box's walls (glued together with Gorilla glue), the inner walls that supported the electronic equipment consisting of two digital recorders, two tie-pin microphones and a small portable speaker (see Fig.1) The choice of equipment was based on the light weight requirements for our payload. To fly on a balloon, the total weight of the payload needed to be on the order of 1 kg.
\\The box had a length of 62 cm, a width of 16 cm and a height of 7.5 cm.
The box material was made of Elmer's foamboard (with a thickness of about 0.5 cm), that was chosen for its light weight, thermal insulation properties and affordable price.
We have implemented a design for the box with one the of the sides much longer than the others. This particular design (different from the typical LA-ACES cube box) was meant to facilitate a longer arrival time delay between the microphones, thus improving our measurement's precision and accuracy. Inside the box there were two compartments of which, one was used to store the player, the recorder and the batteries, and the other one was filled with insulating materials. The box was wrapped in aluminum tape to improve its insulating properties and to give more structural strength to the payload. The payload was provided with PVC straws, glued to the inner walls, so that it could be attached to the balloon by strings.
The recorders were safely stored inside the payload box and insulated from the external environment with fiber glass wool, that provides both good padding and thermal insulation. 
Two small tie-pin microphones (model Sony ECMC 115) were placed, partially externally, along the axis of one of the box's longer sides, and situated at a distance of 33 cm from each other. The two microphones cables were first spliced, and then connected to a single audio cable that was protected by insulating material inside the box, as the right and left channels inputs. The audio cable then was connected, as a single stereo audio input, to one of the recorders. This recorder was our main data acquisition unit (model Olympus Linear PCM Recorder LS-10).  
In addition to recording the environmental sounds, we produced a sinusoidal signal, at a frequency of 1920 Hz, to measure the delay in arrival time between the two microphones. The signal was sampled at a sampling rate of 96kHz (this is the fastest commercially available sampling rate in portable recorders). A high sampling rate and equivalently a short sampling time, compared with our typical millisecond time delay, is necessary to guarantee a high accuracy in the measurement of the speed of the sound. The signal was played by one of the recorders (used as a player also with a sampling rate of 96kHz) and fed to a portable speaker (model Sony T30).
The portable speaker was attached to one of the sides of the payload box at a distance of 26.5 cm from one of the microphones. 
This is a distance bigger than the beginning of the theoretical radiation zone, situated at 18 cm from the source (this last distance is about one wavelengths for a 1920 Hz signal). We tested our equipment for the shortest distance that would give acceptable results and we have determined that the chosen distance is a good compromise between being in the radiation zone, sound level at the microphones and a reasonable size for the payload box.\\
The microphones and speaker were aligned, as accurately as possible. The speaker was wrapped in aluminum tape to insulate it from temperature changes; then it was safely attached to the payload box with cable ties and many layers of Scotch tape and aluminum tape. We also provided some padding between the speaker and the box (to minimize vibrations) through some packing material (we avoided bubble paper that can cause explosions at high altitude). Each recorder was powered by 2 AAA batteries and the speaker by 3 AA batteries. The speaker had a total power of 2 Watts. Sound absorbing material was placed between the microphones to reduce interference from waves reflected from the walls of the box.\\
Usually electronic equipment needs to operate within temperatures that are close to room temperature. One of the main technical challenges of the LA-ACES project is to deal with the extreme temperatures of the upper atmosphere, which can reach minus 50 degrees Celsius. We decided to rely on passive methods of insulation (such as fiber glass wool), the sturdiness of the equipment and the heat generated by the speaker and recorders' batteries to keep the equipment within operational temperatures.
We have tested our payload box in a freezer that had an average temperature of minus 80 degrees Celsius for about 35 minutes and everything worked perfectly. As expected, our payload equipment survived the harsh environment of near space.

\section{Principle of operation} 	
The main principle of operation of our apparatus is based on the simple idea that the sound generated by the speaker travels at a given speed (that is mainly temperature and therefore altitude dependent) and reaches the two microphones at two different times. Therefore, a precise delay is expected between the arrival times of the sound wave at the two microphones. The speed of sound can be determined, then, by dividing the known distance between the two microphones by the time delay. We measured the time delay at each signal cycle, using the arrival time at the peak of the sine wave for left and right channels, and took averages at every quarter of a second. These averages were our experiment's main time sequence, see Figure 2. The speed of sound was measured as function of time at different altitudes. The time in our figures is counted from the the beginning  of our data recording that started 36.5 minutes before take-off. The LA-ACES organization provided altitude data as a function of time acquired through the GPS system that was carried in one of the payloads of the balloon payload train. We compared this altitude time sequence with our speed of sound time sequence. Synchronization between the speed of sound time sequence and the GPS data was aided by the recording of the voice of the flight director that announced verbally the precise moment of the launch through the command "Launch!" 
\section{Software Design and Implementation}
Our main reference signal was produced using a MatLab code that generated a .wav sound file at a given sampling frequency. The .wav file (two minutes long) was stored inside one of the recorders that was set on song repeat to generate a continuous signal that was captured by the other recorder connected to the microphones. The data PCM sound file was analyzed using a well tested MatLab code that the students have written. First, the code read the data from the stereo audio file that had a 96kHz sampling rate and a 16 bit per second memory format. At this rate, we collected 2 Gigabites of data during the first hour of the flight (up to the beginning of the Stratosphere). Then the data from the two channels were stored as two separate time variables. A band pass Chebychev filter, around the signal frequency (with a bandwidth of 50 Hz), was applied to the data to reduce environmental noise that can cause errors in the determination of the time delay. The filter also eliminated low frequency noise due to the swinging motion (common in a balloon flight) of the payload. The code finally divided the data from the two channels in intervals that had a time duration equal to the signal period. The time for the maximum of each signal cycle and each channel was recorded and the time difference between the maxima was measured. The known distance between the two microphones was divided by the time difference to give us the speed of sound as a function of time. 
\section{Data Analysis}
Considering the simplicity of our design, the light weight of the experimental device, the considerable environmental noise and the motion of the payload, our experiment showed a remarkable accuracy and precision. In the laboratory we have compared the measured speed of sound with the known speed of sound for a given temperature. The data exhibited a Gaussian distribution with an average slightly away (about 5 percent in percentage difference) from the accepted value and a 0.7 percent standard deviation. The difference between accepted value and experimental value was interpreted as a calibration issue (mainly due to assumed perfect alignment between microphones and speaker). In our successive calculations, we adjusted the known distance between the microphones to take into account this discrepancy.\\
During the flight we experienced substantial environmental noise, mainly due to the motion of the balloon and the presence of gushes of wind.
Our experimental set up proved very reliable even under such circumstances.
Given the precision of our measurements, an important consideration was to correct the speed of sound by taking into account the average ascension velocity of the balloon. The balloon velocity was calculated from the altitude data (as a function of time) given by the GPS transmitter.  
We calculated the percentage difference between the experimental data and the values of the speed of sound based on a theoretical model that assumes a medium made of an ideal gas with the molecular weight of air. The temperature data used to compute the model were acquired by a temperature probe in the GPS payload attached to the balloon. Figure 4 illustrates the percentage difference values. The agreement between theory and experiment is excellent, in particular in the region of the Troposphere (the first segment of our data set). The average percentage difference in this region is about 0.2 percent while the standard deviation is about 0.7 percent of the measured speed of sound. The transition from the Troposphere to the Tropopause and from the Tropopause to the Stratosphere is marked by fast changes in the balloon ascension velocity. These changes are clearly visible in our data (see Figure 2 and 3) as sudden changes in the trend of the speed of sound (what was measured was actually the speed of sound plus the balloon upward velocity). Because the altitude GPS data were collected with a too low sampling rate (once every minute), we could not obtain a good fit between theory and experiment in the regions of the Tropopause and Stratosphere, even after adjusting for the balloon velocity. The percentage difference between experimental and accepted values is yet very small, even in these regions (about 2 percent).

	\begin{figure}
	\centering	\includegraphics[width=0.45\textwidth]{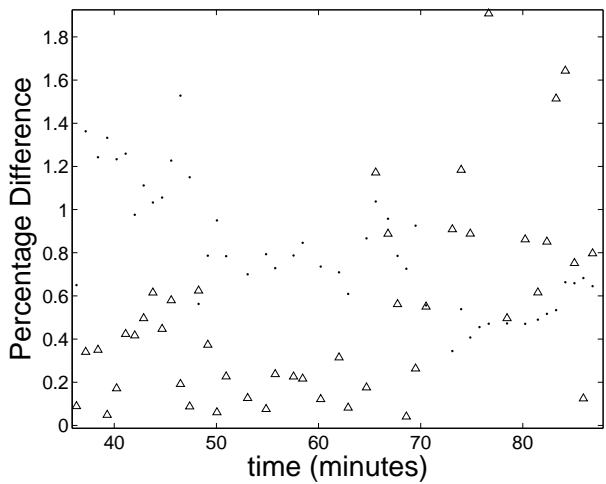}
	\caption{This graph shows the remarkable accuracy and precision of our experiment.  The triangles are the percentage difference between the theoretical and the experimental values of the speed of sound (in the region of the Troposphere the average value of the percentage difference is about 0.2 percent). The first data point, excluded in the above figure, are off from the theoretical value because they represent data collected on the ground when there was considerable amount of high frequency noise, caused by human activity. Yet, the difference between theory and experiment is at most 2 percent. Once the payload starts to fly (at minute 36.5 from the initial recording time), the noise reduces by a factor of about 4. The difference between the theory and experiment increases again in the regions of the Tropopause and Stratosphere for possible reasons explained in section VI. This is the largest discrepancy in our data during the flight and again the percentage difference is less than 2 percent. The dots represent the percentage of the values of the standard deviation with respect to the measured values of the speed of sound. In average the percentage of the error is about 0.7. }
	\label{fig:PE1}
\end{figure}

\section{Estimation of limits on the abundance of Nitrogen in the Atmosphere using the speed of sound}
While the physics involved in the measurement of the speed of sound is quite simple, the dependence of the speed of sound on different parameters is much more interesting \cite{c3}. For a gas in general, the speed of sound $c$ is given by:
\begin{equation}
c=\sqrt{\gamma\;\frac{p}{\rho}},
\end{equation}
where $\gamma$ is called the adiabatic index, $p$ is the gas pressure and $\rho$ its density. The important constant $\gamma$ is the ratio of the specific heats of a gas at constant pressure to a gas at constant volume.
For an ideal gas, we can use the equation of state $p=nRT/V$ together with $\rho=nM/V$, thus:
\begin{equation}
c=\sqrt{\gamma\;\frac{R}{M_{gas}}\;T}=\sqrt{\gamma\;\frac{R}{M_{gas}}\;\left(\vartheta+273.15\right)}
\end{equation}
where $M_{gas}$ is the gas' molar mass in kilograms per mole, the molar constant $R=8.315$ J $mol^{-1}\: k^{-1}$, T is the absolute temperature in kelvins and $\vartheta$ is the temperature in degrees Celsius. The value of $\gamma$ depends on subtle atomic properties of the gas, but a value of 1.4 is a good approximation for diatomic molecules as nitrogen and oxygen that are the main components of the atmosphere.\\
We have calculated the theoretical speed of sound for different molecular masses of the gas. For simplicity, we assumed a binary composition of the atmosphere, as made principally of nitrogen and oxygen. When the relative abundance of nitrogen to oxygen is changed, the molecular mass of the gas changes slightly and thus the speed of sound. In order to distinguish between different models and set significant limits on the composition of the atmosphere, very accurate measurements of the speed of sound are required. The data of our experiment pass through the curve that represents the realistic composition of the Earth's atmosphere (see Figure 5). It seems, though, that because of the sensitivity of the speed of sound to the abundances of the medium main components, even our relatively small error-bars are not able to set strict limits on the atmospheric composition. In the next section, we use Bayesian methods to tackle the particular problem of quantifing the validity of different theoretical models given our data and their uncertainties.

\begin{figure}
	\centering	\includegraphics[width=0.45\textwidth]{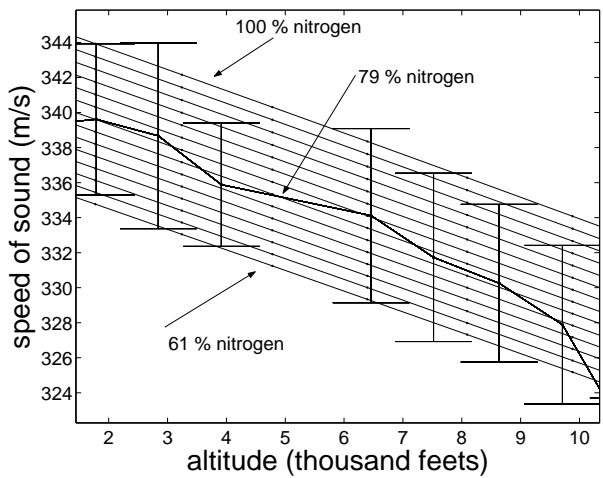}
	\caption{The speed of sound is very sensitive to the composition of the medium. To calculate the theoretical speed of sound we have used a model that implies an ideal gas but with a range of different molecular masses. The different masses are inferred assuming a binary diatomic composition for the gas. The gas is a mixture of oxygen and nitrogen with a successive decrease of nitrogen versus oxygen starting from the top of the figure. The first line from the top is the speed of sound (as a function of altitude) for a gas that is composed exclusively of nitrogen and the last line is that one for a gas with about 61 percent of nitrogen and 39 percent of oxygen. The data points, corresponding to the measured speed of sound in the Troposphere (here, for clarity, we have zoomed in a small range of altitudes), are going through the correct percentage of nitrogen for a realistic atmosphere, that contains about 79 percent of nitrogen and 21 percent of oxygen. Unfortunately, the error bars don't seem to set very strict limits on this particular value of the composition. To better determine the uncertainty on the atmosphere's composition, given by our data, we have used Bayesian methods to set limits on the abundance of the elements (in a way similar to what was done in a recent paper that estimated the composition of the atmosphere of Titan using the measured speed of sound). See Figure 6 and 7.}
	\label{fig:atmosphere1}
\end{figure}

\section{Bayesian Analysis}
\begin{figure}
\centering	\includegraphics[width=0.45\textwidth]{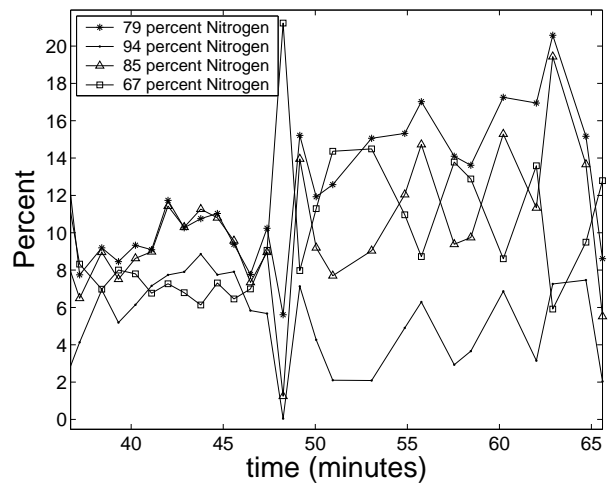}
	\caption{The likelihood values versus time for 4 models of the composition of the atmosphere. The graphs represent the probability of obtaining data $D_{i}$, given the model $M_{i}$ and prior information $I$ being true. The prior information $I$ used is that all the models are similarly plausible at first. The data $D_{i}$ are the measurements of the speed of sound as a function of time. The models used have values of the composition of the atmosphere with 79, 94, 85 and 67 percent nitrogen, respectively.  We have assumed in our Bayesian calculation that the uncertainty in our measurement is represented by a Gaussian distribution with standard deviations equal to our measured sigmas (one for every data point). The most likely model seems to be the one with a realistic composition with 79 percent nitrogen. For the first 15 minutes the models with a composition of 79 and 85 percent are practically indistinguishable; afterwards the model with realistic values of the composition is favored.}
	\label{fig:atmosphere1}
\end{figure}

Bayesian statistic or probability is becoming an increasingly important analysis tool in the physical sciences \cite{c5}, \cite{c8}.
A good reference, accessible to undergraduate students, is Gregory \cite{c2}. In particular, Bayesian methods are very useful in comparing two or more competing theoretical models given our state of the knowledge (our data and their uncertainties).
The Bayesian analysis foundation is the Bayes's Theorem:
\begin{equation}
	p(H_{i}|D,I)=\frac{p(H_{i}|I)\:p(D|H_{i},I)}{p(D|I)}
\end{equation}
where $H_{i}$ is an hypothesis we want to test, $I$ is the prior information, $D$ is the data available, $p(D|H_{i},I)$ is the probability  of obtaining our data given $H_{i}$ and $I$, $p(H_{i}|I)$ is the prior probability of our hypothesis, $p(H_{i}|D,I)$ is the posterior probability of the hypothesis (given data and prior information) and finally, $p(D|I)=\sum_{i}p(H_{i}|I)\:p(D|H_{i},I)$ is a normalization factor that guarantees $\sum_{i}=p(H_{i}|D,I)=1$. The quantity $p(D|H_{i},I)$ is also called the likelihood.
\\We can use the Bayes' theorem to test the validity of different theoretical models. Model comparison is based on the specification of two or more alternative models.
We use the prior information that one of the models is true and that the data set a Gaussian distribution with a certain measured standard deviation. This is exactly our case. Therefore, we have:
\begin{equation}
p(Mi|D,I)=\frac{p(Mi|I)p(D|Mi,I)}{p(D|I)}
\end{equation}
where $M_{1}$ is a model based on equation (1) that assumes a molecular composition made of diatomic molecules of nitrogen (with a  realistic relative abundance of 79 percent) and oxygen (comprising 21 percent of the gas). Other models with various percentage combination of nitrogen and oxygen are represented by $M_{i}$ with $i\neq1$.
The uncertainty, e, in speed of sound measurements, is characterized by a Gaussian distribution of the form:
\begin{equation}
p(e|I)=\frac{1}{\sqrt{2\pi}\sigma}\:exp\left(-\frac{e^{2}}{2\sigma^{2}}\right)
\end{equation}
where $\sigma$ is the measured standard deviation of each data point.
We are interested in comparing two models at the time. The two models are the one we assume to be true (with a known nitrogen abundance of 79 percent nitrogen) and a similar model based also on equation (1) but with a different molecular composition (similar in type of atoms present but with different relative percentage of nitrogen). A quantity that is very useful in comparing the two models is the odds ratio, equal to the ratio of the posterior probabilities of the two models. The odds of model $M_{1}$ over model $M_{i}$ are given by:
\begin{equation}
O_{12}=\frac{p(M_{1},D|I)}{p(M_{i}|D,I)}=\frac{p(M_{1}|I)\:p(D|M_{1},I)}{p(M_{i}|I)\:p(D|M_{i},I)}=\frac{p(D|M_{1},I)}{p(D|M_{i},I)}
\end{equation}

\begin{figure}
\centering	\includegraphics[width=0.45\textwidth]{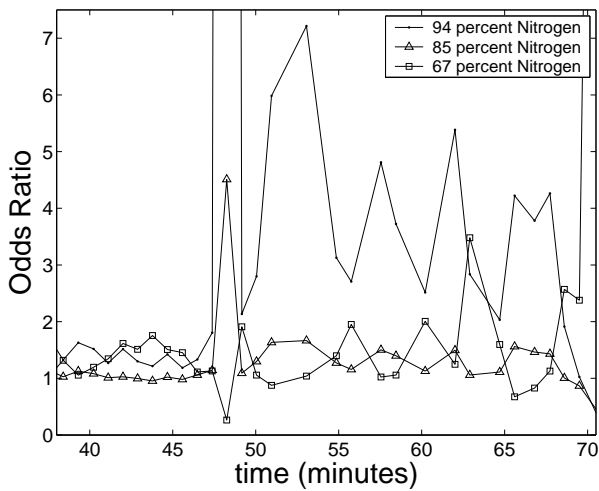}
\caption{Odds ratio for different models of the composition of the atmosphere.  The graphs represent the function $O_{12}$ that is the ratio of the posterior probabilities of model $M_{1}$ over $M_{i}$. The model $M_{1}$ is assumed to be true (this is our known composition model with 79 percent nitrogen). The model $M_{i}$ correspond to different compositions of nitrogen with relative abundances of 94, 85 and 67 percent. The odds ratios are not very high for our favored model in the first 15 minutes but become better in the remaining of the flight (in the figure we show just the region of the Troposphere). Testing different models, we make a relatively safe bet (with average odds of about 4 to 1 over the entire flight) that the composition of the atmosphere, given by our data, is in the range between 85 and 73 percent nitrogen and the rest oxygen (so our estimated range is within 8 percent of the accepted value). }
	\label{fig:atmosphere1}
\end{figure}

The quantity $p(M_{1}|I)/p(M_{i},I)$ is equal to 1 because our prior information tells us that both models are equally valid at first.\\
In order to calculate the likelihood $p(D|M_{i},I)$, we make the assumption that the model $M_{1}$ is true. Therefore, we justify the discrepancy between our speed of sound measurements $v_{i}$ and $v_{M_{1}}$, the speed given by the true model, because of random errors in the measurements. Then, it follows that $v_{i}=v_{M_{1}}+e$ or $
e=v_{i}-v_{M_{1}}$. The probability of obtaining the measured data, given our favored model and our prior information being true, can be obtained by:
\begin{equation}
p(D|M1,I)=\frac{1}{\sqrt{2\pi}\sigma}\:exp\left(-\frac{\left(v_{i}-v_{M_{1}}\right)^{2}}{2\sigma^{2}}\right)
\end{equation}
The likelihood values as a function of time, for different models, are illustrated in Fig. 6.
The odds ratio of selected models, compared to the realistic composition model, are shown in Fig. 7.
It is straightforward to derive the following result, the posterior probability of each model, compared to the true model, is given by:
\begin{equation}
p(M_{i}|D,I)=\frac{O_{i1}}{1+O_{i1}}.
\end{equation}
For example, if we compare the model with 79 percent nitrogen and the model with 85 percent nitrogen we obtain a posterior probability  (over the entire flight) of 80 percent for the former and 20 percent for the latter. The model with 67 percent nitrogen is slightly more favored than the one with 85 percent (even if the two models are equally separated from the true model). This is due to the fact that our data points are, in average, below the realistic model curve. Models with relative abundances closer to realistic values have posterior probabilities nearly on the order of the one of the true model. This means that the odds to distinguish between the real model and a slightly different one become too small. In other words is not possible to set significant limits using models with nitrogen percentage below 85 (or above 67) percent. Given our data, we can, then safely (with odds of about 4 to 1) conclude that the atmosphere's composition has an abundance of nitrogen between 73 and 85 percent, or within 8 percent of the accepted value.

\section{Conclusions}
We have shown a simple design for the measurement of the speed of sound that is very portable, robust and simple. The proposed speed of sound apparatus can be used in the laboratory or in the open (we used it as a payload on a high altitude balloon). Our experimental set up allowed us to obtain atmospheric measurements with great accuracy and precision even under extreme conditions of low temperature, pressure and abundant environmental noise.\\
We were able to use our speed of sound measurements, through the application of Bayesian analysis, to set significant limits, in particular for a student project, on the atmospheric composition of the Earth. The composition is between 85 percent and 67 percent nitrogen (with odds larger than 4 to 1) and the rest oxygen.\\

\begin{center}
\textbf{\small AKNOWLEDGEMENTS}
\end{center}
This work was supported by LA-Space grant number 0000000.\\
We would like to thank Dr. Guzik, Dr. Wefel and Mr. Giammanco of Louisiana State University for their support and assistance. Many thanks to the NASA employees at the Columbia Scientific Balloon Facility, in Palestine, Tx, for their patience and time.


\bibliographystyle{unsrt}
\bibliography{SpeedofSound2}
\end{document}